# PREY-PREDATOR MODELING OF $CO_2$ ATMOSPHERIC CONCENTRATION

LUIS AUGUSTO TREVISAN

*Departamento de Matemática e Estatística, Universidade Estadual de Ponta Grossa, Avenida Carlos Cavalcante*

*4748*

*Ponta Grossa, Paraná, ZIP 84030-000, Brazil.- Correspondin Author: luisaugustotrevisan@yahoo.com.br*

FABIANO MEIRA DE MOURA LUZ

*Departamento de Física, Universidade Estadual de Ponta Grossa, Avenida Carlos Cavalcante 4748*

*Ponta Grossa, Paraná, ZIP 84030-000, Brazil.*

In this work we propose a mathematical model, based in a modified version of the Lotka-Volterra prey-predator equations, to predict the increasing in $CO_2$ atmospheric concentration. We consider how the photosynthesis rate has changed with the increase of $CO_2$ and how this affects plant reproduction and $CO_2$ absorptions rates. Total CO2 emissions (natural and manmade) and biomass numerical parameter changes are considered. It is shown that the atmospheric system can be in equilibrium under some specific conditions, and also some comparisons with historical are done.

# 1. Introduction

Several billion of years ago, special terrestrial conditions made the formation of life possible. Since its formation, the biosphere has played an active role in controlling environmental conditions. An interaction has developed between the evolution of living species and the environment. Changes in environmental conditions modified the biosphere and vice-versa. When life began, the Earth's atmosphere was not similar to the present air. The main peculiarity of our atmosphere - the presence of oxygen – is the result of the biospheres evolution (Meszaros). Natural changes in atmospheric composition and climate are slow processes when compared with typical human time scales. During the last 8.000 to 10.000 years, the climate has been stable. Such stability has been favorable for humans and made social and economic development possible. In the present industrial era, this development has reached such a level that human activities have become able to modify environmental conditions on a time scale ($\approx$ 100 years) that is quite shorter than periods of natural changes. It is well known that the $CO_2$ emission due to human activities has contributed to cause the "greenhouse effect", a warning in the Earth mean temperature. Another important effect of the increase of the $CO_2$ concentration is acceleration in the photosynthesis rate, a subject widely studied (Kirsshbaum)(Laisk and Edwards). With a higher photosynthesis rate, plants can absorb more $CO_2$, and have a faster growing and reproduction rate, again consuming more $CO_2$. In this work we present a model to predict the increase of $CO_2$ atmospheric concentration, considering the plant-atmospheric carbon interaction. Basically, we consider the increase of photosynthesis rates, that is a consequence of the increase of $C_a$ ($CO_2$ atmospheric concentration), and apply a modified version of Lotka-Volterra predator-prey model to describe a possible time evolution of $C_a$ and a numerical biomass parameter. This parameter includes all life beings that use photosynthesis, even if they are not vegetables.

2.  **Predator-prey model adapted to plant-atmospheric carbon interaction**

The original model by Lotka-Volterra (Lotka,Volterra) was proposed in the twenty's, and have been widely used in ecological studies and also in another fields of knowledge. In our model, adapted to the problem of plants $CO_2$ interaction, we will consider the $CO_2$ as the inorganic prey (not reproducing) and plants are predator. The $CO_2$ molecule does not reproduce by itself, therefore its concentration evolution, denoted by $C_a$ (in ppm), depends only on emissions (i.e. combustion reactions) denoted by Q(t), and absorptions (i.e photosynthesis), assumed to be proportional to vegetal biomass surface, denoted by P(t) and also to the relative photosynthesis rate $A(C_a, T)$, where T is the temperature. In this way, we have the set of equations (*k* is a constant):

$$\frac{dC_a}{dt} = -A(C_a,T)*P + kQ(t)$$
$$\frac{dP}{dt} = -e*P + f*P*A(C_a,T)$$

(I)

The relative (in comparison with the maximum possible) photosynthesis rate is given, at temperature of 25°C by (Kirshbaum)

$$A(C_a,T) + \frac{V_j(C_a - 61)}{C_a + 126} \qquad (2)$$

To estimate k, we consider the work by P.Tans, F.Y.Inez and T.Takahashi, that studied emissions and concentration during the 80's, the value is obtained considering the annual average increase on concentration of 1.4ppm and difference between emissions and absorption, which the mean value is 3GtC/year. The ratio reads k= 0.47. We also consider that the oceans are a balanced system, that absorbs all the $CO_2$ they emit and is also able to absorb around 30% of human emission (Sabine et.al). Equation 1 says that the growing of the numerical value for P is proportional to itself and also to photosynthesis rate. The last equation is rewritten

$$\frac{dP}{dt} = f * P(-b + A(C_a,T)) \qquad (3)$$

Where $b = e/f$, $b \geq 0$. The total $CO_2$, $Q(t)$ emission can be separate in natural emissions $Q_n$ and human emission $Q_h(t)$.

$$Q(t) = Q_n + Q_h(t) \qquad (4)$$

We have important remarks about the balance in the system. First, we note that if the system is balanced, $b = A(C_a, T)$, the photosynthesis rate is equal to vegetable death rate (b factor). Then, the first balance condition is

$$b < 1 \qquad (5)$$

This fact is explored in this work, to show how we can use the parameter *b* to study possible stabilization scenarios in the atmospheric $CO_2$ concentration. When:

$$dC_a/dt < 0 \qquad (6)$$

A becomes smaller while $dP/dt$ increases, and as time increases $|dC_a/dt|$ becomes smaller, approaching zero and a final balance condition. On the other hand, if

$$dC_a/dt > 0 \qquad (7)$$

a necessary condition to reach balance conditions is:

$$d(A*P)/dt > dq(t)/dt \qquad (8)$$

that is, the absorptions must increase faster than emissions, until a balanced system ($dP/dt = 0$ and $dC_a/dt = 0$) is reached.

The b factor is time dependent (more exactly, depends on the historical moment) and is also related to the human emissions, since some fraction of the vegetable death and human emission are due to the same fact: burning. Deforesting and urbanization also contribute to increase the b factor. In this way

$$b \rightarrow b(t) \qquad (9)$$

Besides, the f value can be used to help to fit desirable value of the derivative.

In this way, the model has only two free parameters, $b(t)$ and $f$ that could be estimate using environmental data.

To better understand the role of the *b* factor, we should observe figures 1 and 2 below: In figure 1, we use the same human emissions ($Q_h = 7.5GtC/Y$, which taking into account ocean absorption, should be read 0.7x7.5=5,25GtC/Y). In cases 1, 3 and 5 we have used the same *f* value ($f = 0.0185$), the value of the *b* factor is supposed constant and fitted to given three different final CO2 concentrations; 500*ppm*; 600*ppm* and 700*ppm*, corresponding to *b*=0.70; b=0.74 and b= 0.77 respectively. The initial concentration is $C_a = 345ppm$. Therefore there is a relantionship between the vegetal death rate (b) and the end carbon dioxid atmospheric concentration:

$$b = A(C_a, end), 25) \qquad (10)$$

Besides this, in curves 2 and 4, we have used the same *b* value of curve 3, b=0.74, with different values for f ( f= 0.03 in curve 2 and f=0.01 in curve 4, while f=0.0185 in curve 3). We observe that, using the values f=0.0185, and b=0.74 (which corresponds to final concentration of 600ppm) we may fit the data between 1985 (345ppm, jan) [8] and 1995 (360ppm, jan ) .

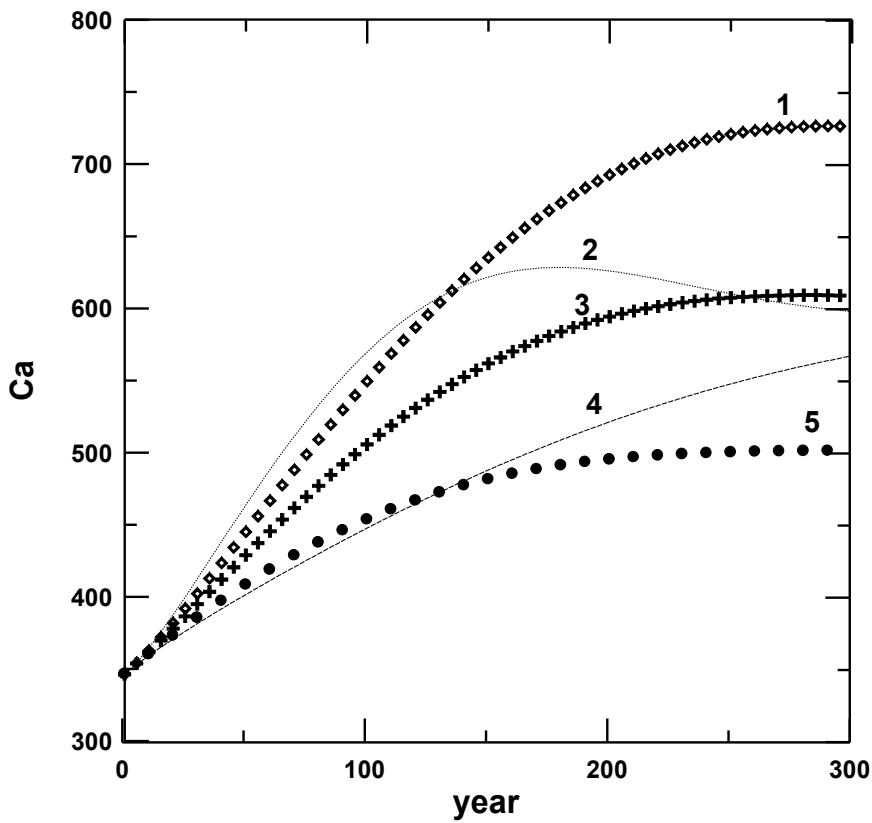

Figure 1. $CO_2$ concentration versus time for different b and f constant values. Curve 1: b=0.77, f=0.0185; curve 2: b=0.74, f=0.03, curve 3:b=0.74, f=0.0185, curve 4: b=0.74, f=0.01; curve 5 :b=0.70, f=0.0185.

In figure 2 below, we used the same constant $b$ from figure 1 and show the time evolution of P/P0, being P0 the initial value. We also observe stabilization, after some decreasing period.

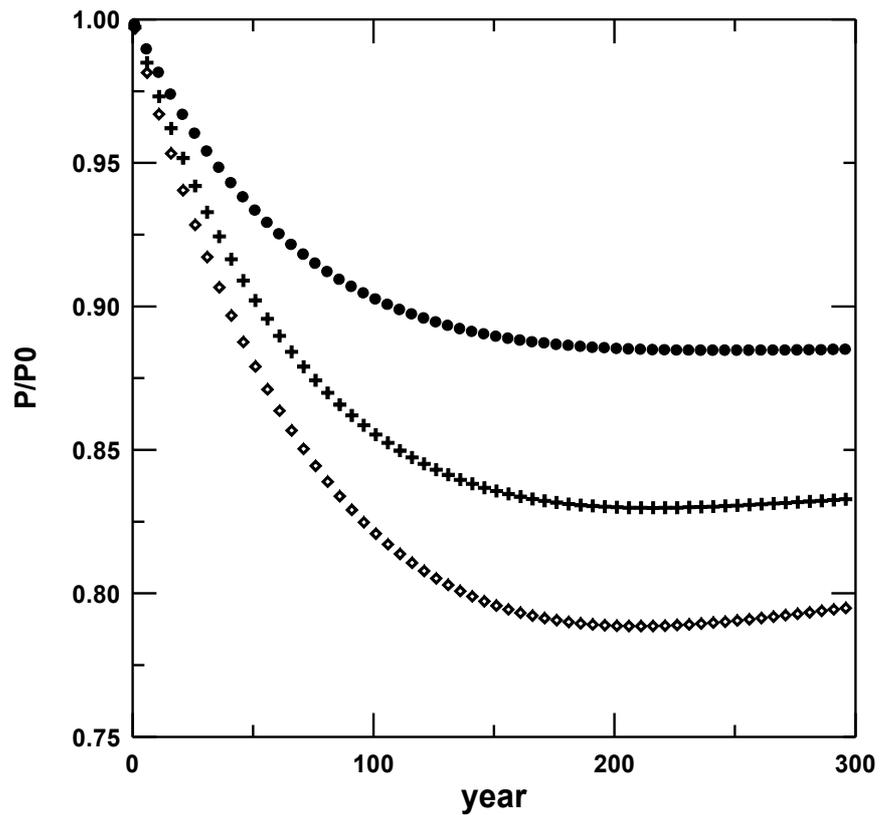

Figure 2. Normalized biomass parameter evolution, for different constant values of the death rate $b$. Circles mean b=0.70, crosses mean b=0.74 and squares mean b=0.77.

We emphasize that the main difference of the present model to the IPCC is the way the absorptions are considered The system can reach a stable concentration even with constant emissions when the *b* factor is less than unity (the greatest value the relative photosynthesis rate can reach)

3. **Results**

The model parameters are adjusted to give results near the data of $CO_2$ concentration in the 1800-2000a.c. period, when the concentration increased from approximately 282 ppm (Neftel et al) up to 370 ppm (2000), and with the recent annual average increase of 1.4 ppm ( in the 80's)(Tans,Ynez,Takahashi). We also suppose the human emissions ($Q_h$) to have an exponential growing, from nearly zero up to the present level. We have used the value 7.5 GtC/year (1985). We note that the best result is obtained using the *b* factor growing exponentially with time. The initial *b* value is: $b_i = A(282\,ppm; 25^\circ C)$, that is, the initial *b* factor was equal to photosynthesis rate because, at that time, the system has been supposed to be balanced. And the final *b* value, combined with emissions, gives the increase of concentration that is close to 1.4 ppm/year, at 80's and a concentration near the present one. Actually, we use the following time functions to emissions:

$$Q_h = \exp(\ln(750)*t/185)*0.01 \tag{14}$$

where $Q_h$ is the human emission, and t is the time, in years. The value t = 185 is equivalent to 1985, and correspondingly $Q_h$ = 7.5GtC/y. For the *b* factor we used:

$$b = b_i * \exp[t * \ln(0.74/b_i)/185] \qquad (15)$$

so, if t = 185, b = 0.74. As an important information, in the model, the natural emission ($Q_n$) and absorption of $CO_2$ are initially supposed equal, 100GtC/year (the same of nowadays), where we consider only terrestrial absorptions and emissions. The human absorption is 2.25GtC/year (1985), close to zero at 1800, and we consider that absorptions due the oceans are around 30% of human emissions, so we multiply the human emissions by 0.7. The initial value for the P parameter is fixed by:

$$\frac{dC_a}{dt} = k(Q_n + 0.7Q_h) - A(282, 25) * P = 0 \qquad (16)$$

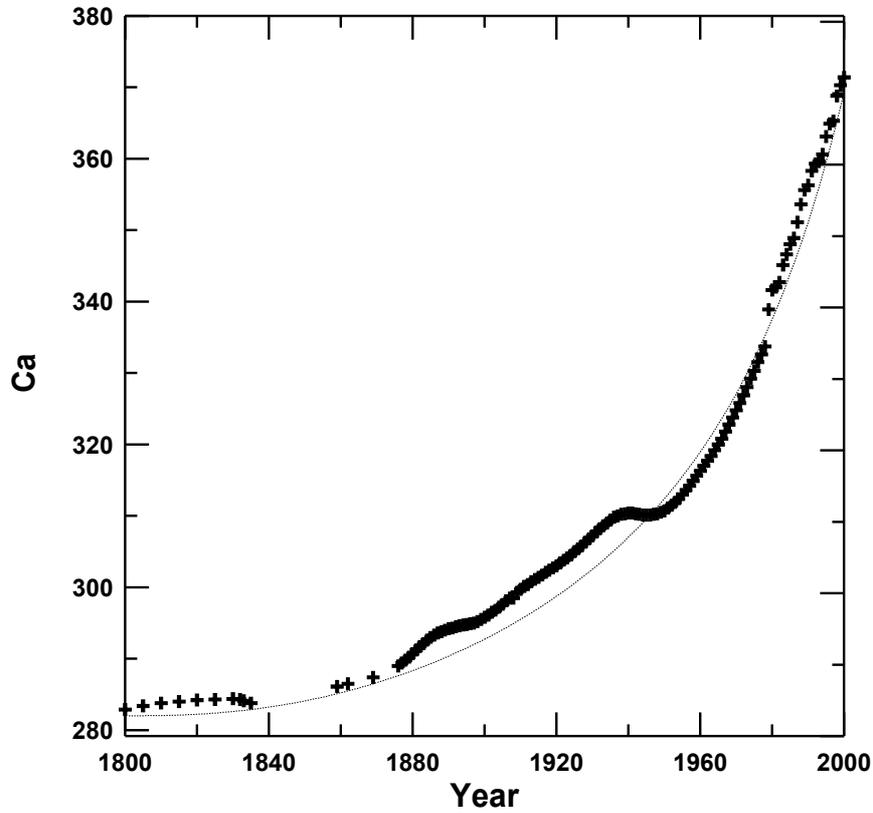

Figure 3 Comparison between the data from (Keelin et al) (Wahlen et al) (astries) and the fit of the model (dashed line)

Using $f$ = 0.0055 and the equations above we got the fitting shown in figure 3 (dashed line) and extrapolated them to obtain table 1, with predictions of $C_a$ and its annual increase in 2010 and 2015. In table 1 the historical data are compared with the results arising from our modell, in 1995 and 2000a.c., respectively.

In the figure 4 below, we fit a short time evolution for the concentration, considering constant emissions (7 GtC/y), ocean absorption around 30% of human emission, vegetal death rate 0.78 and f = 0.015

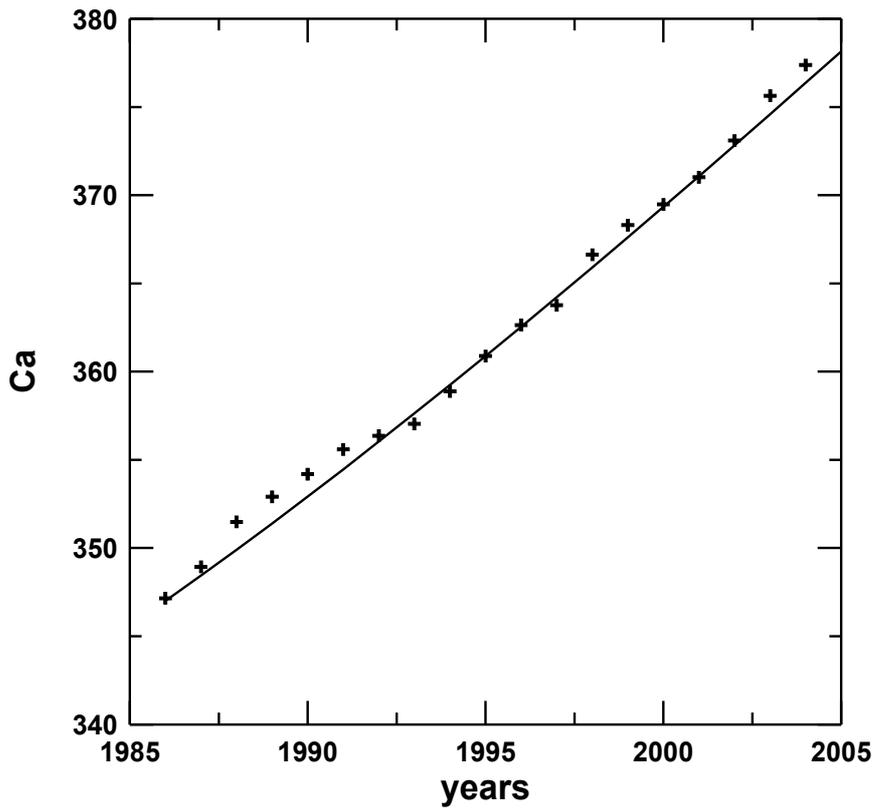

Fig 4: Concentration in a short time scale. We notice an almost linear evolution.

Table 1: $C_a$ predictions, according to the present model and historical data

| After High year | Prediction | | Historical data | |
|---|---|---|---|---|
| | $C_a$(ppm) | $\frac{dC_a}{dt}(ppm/y)$ | $C_a$(ppm) | $\frac{dC_a}{dt}(ppm/y)$ |
| 1995 | 359.54 | 1.78 | 363.10 | 1.80 |
| 2000 | 369.34 | 2.08 | 371.40 | 2.20 |
| 2005 | 380.82 | 2.45 | 377.43-dec/04 | |
| 2010 | 394.38 | | | |
| 2015 | 410.48 | | | |

The dashed line in Fig. 3 corresponds to an exponential increase in the emissions, takes into account a variation in the absorptions due the increase of the b factor, and also the increase of the photosynthesis rate. So with the b factor being "`time dependent"', growing exponentially from zero to the present value , constant *f* parameter and with simulated exponential emissions (not free, they should comes from historical data) we obtain a smooth fit to historical data. We must note, in the historical data, the period 1938-1946, when a constant value of Ca is observed.. Another important point is that only with raising emissions it is not possible to fit the data curve completely. Only considering variations in absorptions, due the vegetable death, the curve may be reasonably fitted. This means, for example, that burning forest contributes twice, by one side $CO_2$ concentration increases due to emissions (burning) and by the other side it also increases due to reduced absorption (deforesting). As an additional remark about the features of the model, we notice that the parameters to fit short range data may be different to those needed to fit a long historical period, this is intrinsic to the non-linearity of the set of coupled equations.

4. **Conclusion**

In this work we presented a mathematical model, based on prey-predator equations, to estimate the time evolution of the atmospheric $CO_2$ concentration. We consider the effect of a growing photosynthesis rate when $CO2$ atmospheric concentration increases. In order to simplify, we study only one kind of photosynthesis ($C_3$) and the ambient temperature was kept constant. Using the photosynthesis rate function in a adapted Lotka-Volterra system of equations, where $CO_2$ where prey and plants are predators we simulated several scenarios and show that the parameters can be choice to reproduce the historical data and the IPCC predictions. It is important to note that our model has only two free parameters *f* (related to the derivative *dP/dt)* and the most relevant *b*, vegetable death rate. We note that $0 \leq b \leq 1$ and when $A = b$ we have equilibrium. That is, the photosynthesis rate in the equilibrium is equal to vegetable death rate. This is the reason because it is easy to fit the initial an final values in the model. With *f* we fit the derivatives, and with *b* the end point (stabilization). We also observe that the model consider a different way to compute only the absorptions of $CO_2$, keeping the emissions as an external input. As future extension of this study, we intend to consider other effects, like temperature variation, $C_3$ photosynthesis and climate forcing.

**Acknowledgments**

Both authors thank Prof. R. A. Clemente for criticism and suggestions. F.M.M. Luz thanks the State University of Ponta Grossa for financial support.